\documentclass[dvips,fleqn,twoside]{article}
\usepackage[headings]{espcrc2}
\usepackage{epsfig}

\readRCS
$Id: espcrc2.tex,v 1.2 2004/02/24 11:22:11 spepping Exp $
\usepackage{graphicx}
\usepackage[figuresright]{rotating}

\newcommand{\Li}[2]{{\mbox{Li}}_{#1}\left(#2\right)}

\newcommand{\Snp}[2]{{\mbox{S}}_{#1\!}\left(#2\right)}

\newcommand{\be}{\begin{equation}}
\newcommand{\ee}{\end{equation}}
\newcommand{\bea}{\begin{eqnarray}}
\newcommand{\eea}{\end{eqnarray}}
\newcommand{\ep}{\varepsilon}

\newcommand{\tfrac}[2]{{\textstyle{\frac{#1}{#2}}}}

\title{
{$\;$}\vspace*{-40mm}\\
\begin{flushright}
\large{SINP MSU 2005--23/789}\\[2mm]
\large{September 2005}
\end{flushright}
\vspace*{15mm}
Geometrical methods in loop calculations
and the three-point function}

\author{Andrei I. Davydychev
\address[SLB]{Schlumberger, SPC, 
155 Industrial Blvd., Sugar Land, TX 77479, USA}${}^{,}$%
\address[NPIMSU]{Institute for Nuclear Physics, Moscow State University,
119992 Moscow, Russia} 
}
       
\runtitle{Geometrical methods in loop calculations}
\runauthor{A.I. Davydychev}

\begin{document}

\begin{abstract}
A geometrical way to calculate $N$-point Feynman diagrams is reviewed.
As an example, the dimensionally-regulated three-point function 
is considered, including all orders of its $\varepsilon$-expansion.
Analytical continuation to other regions of the kinematical variables
is discussed.
\vspace{1pc}
\end{abstract}

\maketitle

\section{INTRODUCTION}

The analytical structure of the results for $N$-point Feynman diagrams
can be better understood if one employs a geometrical
interpretation of kinematic invariants and other quantities. 
For example, the singularities of the
general three-point function can be described pictorially 
through a tetrahedron constructed out of
the external momenta and internal masses (see Fig.~1a). 
This method can be used to derive Landau equations defining
the positions of possible singularities \cite{Landau} 
(see also in \cite{3pt_sing}). 

In Ref.~\cite{DD-JMP} it was demonstrated how such geometrical
ideas could be used for an analytical calculation of
one-loop $N$-point diagrams. For example, in the three-point
case in $n$ dimensions, the result can be expressed in terms of an integral 
over a spherical (or hyperbolic) triangle, as shown in Fig.~1b,
with a weight function $\cos^{3-n}\theta$, where $\theta$ is
the angular distance between the integration point and the point 0, 
corresponding to the height of the basic tetrahedron
(see in \cite{DD-JMP}). 
This weight function equals 1 for $n=3$
(see also in \cite{Nickel}). For $n=4$, one can get
another representation \cite{OW,Wagner}, in terms of an integral 
over the volume of an asymptotic hyperbolic tetrahedron. 
Here we will discuss the application of the approach 
of Refs.~\cite{DD-JMP,D-ep}
to the three-point function in any dimension~$n$, 
as well as its $\ep$-expansion ($n=4-2\ep$) within
dimensional regularization~\cite{dimreg}. 
\begin{figure}[t]
\refstepcounter{figure}
\label{3pt_fig1}
\addtocounter{figure}{-1}  
\begin{center}
{\epsfxsize=74mm \epsfbox{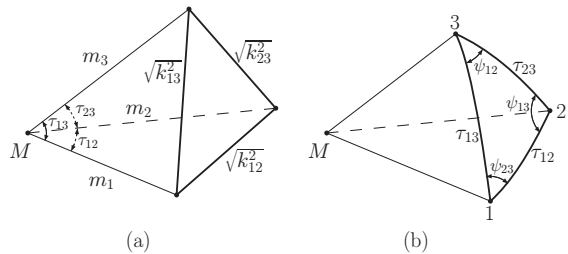}}
\caption{Three-point case: (a) the basic tetrahedron 
and (b) the solid angle}
\end{center}
\end{figure}

\section{TRIGONOMETRIC STUFF}

We will use notations defined in \cite{DD-JMP}, namely:
\begin{equation}
\label{def_c}
c_{jl}=\cos\tau_{jl} 
\equiv \frac{m_j^2 + m_l^2 - k_{jl}^2}{2 m_j m_l} ,
\end{equation}
\begin{equation}
\cos\tau_{0i} = \frac{m_0}{m_i}, \hspace{10mm}
m_0
=m_1 m_2 m_3 \sqrt{\frac{D^{(3)}}{\Lambda^{(3)}}} ,
\end{equation}
\begin{eqnarray}
\Lambda^{(3)} &=& -{\textstyle{1\over4}}
\bigl[ (k_{12}^2)^2 + (k_{13}^2)^2 + (k_{23}^2)^2
\nonumber \\ &&
       - 2 k_{12}^2 k_{13}^2 - 2 k_{12}^2 k_{23}^2
       - 2 k_{13}^2 k_{23}^2
\bigr]
\nonumber \\ 
&=& -{\textstyle{1\over4}} \lambda\left(k_{12}^2,k_{13}^2,k_{23}^2\right) ,
\end{eqnarray}
\begin{eqnarray}
\label{D(3)}  
D^{(3)} &\equiv&
\left|
\begin{array}{c}
{ \; 1\;\;  \;\; c_{12}\; \;\; c_{13}\;  } \\
{ c_{12}\;  \;\; \; 1\;\; \;\; c_{23}\;  } \\
{ c_{13}\;  \;\; c_{23}\; \;\; \; 1\;\;  } \\
\end{array}
\right|
\nonumber \\
&=& 1-c_{12}^2-c_{13}^2-c_{23}^2+2c_{12} c_{13} c_{23} .
\end{eqnarray}
Assuming that all $|c_{jl}|\leq 1$, $\Lambda^{(3)}>0$, and 
$D^{(3)}>0$, we get spherical triangles. 
In other cases, we need to use the hyperbolic space.
The transition corresponds to the analytic continuation
(see below).

Using the approach of Ref.~\cite{DD-JMP}, we split the spherical
triangle into three smaller ones (see Fig.~2), 
\begin{figure}[t]
\refstepcounter{figure}
\label{3pt_fig2}
\addtocounter{figure}{-1}  
\begin{center}
{\epsfysize=36mm \epsfbox{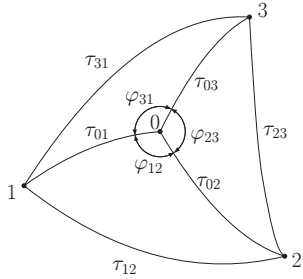}}
\caption{The spherical triangle 123}
\end{center}
\end{figure}
so that 
$\varphi_{12}+\varphi_{23}+\varphi_{31}=2\pi$.
Moreover, it is convenient to split each of the resulting 
triangles into two rectangular ones, as shown in Fig.~3.
By definition,
${\textstyle{1\over2}}\left(\varphi_{12}^{+}
     +\varphi_{12}^{-}\right)=\varphi_{12}$,
${\textstyle{1\over2}}\left(\tau_{12}^{+}+\tau_{12}^{-}\right)
=\tau_{12}$.
\begin{figure}[b]
\refstepcounter{figure}
\label{3pt_fig3}
\addtocounter{figure}{-1}  
\begin{center}
{\epsfysize=36mm \epsfbox{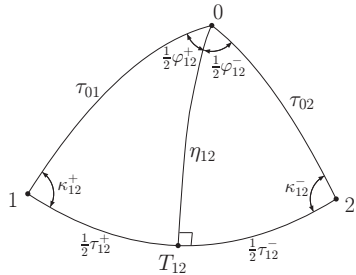}}
\caption{An asymmetric spherical triangle 012}
\end{center}
\end{figure}
Let us list useful relations
between the sides and angles of the spherical
triangle (see also in \cite{DD-JMP}):
\begin{eqnarray*}
\cos\left({\textstyle{1\over2}}\tau_{12}^{+}\right)
&=& \frac{\cos\tau_{01}}{\cos\eta_{12}} ,
\\
\cos\left({\textstyle{1\over2}}\tau_{12}^{-}\right)
&=& \frac{\cos\tau_{02}}{\cos\eta_{12}} ,
\hspace{20mm}
\\  
\cos\left({\textstyle{1\over2}}\varphi_{12}^{+}\right)
&=& \frac{\tan\eta_{12}}{\tan\tau_{01}} ,
\\
\cos\left({\textstyle{1\over2}}\varphi_{12}^{-}\right)
&=& \frac{\tan\eta_{12}}{\tan\tau_{02}} ,
\\ 
\sin\left({\textstyle{1\over2}}\tau_{12}^{+}\right)
&=& \sin\tau_{01} \sin\left({\textstyle{1\over2}}\varphi_{12}^{+}\right) ,
\\
\sin\left({\textstyle{1\over2}}\tau_{12}^{-}\right)
&=& \sin\tau_{02} \sin\left({\textstyle{1\over2}}\varphi_{12}^{-}\right) ,
\\ 
\tan\left({\textstyle{1\over2}}\tau_{12}^{\pm}\right)
&=& \sin\eta_{12} \tan\left({\textstyle{1\over2}}\varphi_{12}^{\pm}\right) ,
\\ 
%\tan\left({\textstyle{1\over2}}\tau_{12}^{-}\right)
%&=& \sin\eta_{12} \tan\left({\textstyle{1\over2}}\varphi_{12}^{-}\right) ,
%\\ 
\sin\eta_{12} &=&  \sin\tau_{01} \sin\kappa_{12}^{+}
= \sin\tau_{02} \sin\kappa_{12}^{-},
\\ 
\cos\kappa_{12}^{\pm} &=&
\cos\eta_{12}
\sin\left({\textstyle{1\over2}}\varphi_{12}^{\pm}\right) . 
\end{eqnarray*}
Worth noting is
\begin{equation}
\cos\eta_{12} = \frac{m_0 \sqrt{k_{12}^2}}
                     {m_1 m_2 \sin\tau_{12}} .
\end{equation}
Therefore,
\begin{eqnarray*}
\cos\left({\textstyle{1\over2}}\tau_{12}^{+}\right)
&=& \frac{m_2 \sin\tau_{12}}{\sqrt{k_{12}^2}}, 
\\
\sin\left({\textstyle{1\over2}}\tau_{12}^{+}\right)
&=& \frac{m_1^2-m_2^2+k_{12}^2}{2 m_1 \sqrt{k_{12}^2}}, 
\\
\cos\left({\textstyle{1\over2}}\tau_{12}^{-}\right)
&=& \frac{m_1 \sin\tau_{12}}{\sqrt{k_{12}^2}}, 
\\
\sin\left({\textstyle{1\over2}}\tau_{12}^{-}\right)
&=& \frac{m_2^2-m_1^2+k_{12}^2}{2 m_2 \sqrt{k_{12}^2}} .
\end{eqnarray*}
In a triangle with the sides $m_1$, $m_2$ and 
$\sqrt{k_{12}^2}$, the angles 
${\textstyle{1\over2}}\tau_{12}^{+}$
and ${\textstyle{1\over2}}\tau_{12}^{-}$
are those between the height of the triangle and
the sides $m_1$ and $m_2$, respectively. 
Therefore, the point $T_{12}$ in Fig.~3 corresponds
to the intersection of this face height and the sphere.

\section{HYPERGEOMETRIC STUFF}

Eqs.~(3.38)--(3.39) of \cite{DD-JMP} yield
\begin{equation}
J^{(3)}(n;1,1,1) = 
- \frac{{\rm i}\pi^{n/2}\Gamma\left(3-\frac{n}{2}\right)}
       {m_0^{4-n}\sqrt{\Lambda^{(3)}}}\;
\Omega^{(3;n)} \; ,
\end{equation}
where $\Omega^{(3;n)}$ is an integral over the solid angle
$\Omega^{(3)}$ (corresponding to triangle 123 in Fig.~1b),
\begin{equation}
\Omega^{(3;n)} = \int\int\limits_{\hspace*{-5mm}\Omega^{(3)}}^{}
\frac{\sin^{n-2}\theta \; {\rm d}\theta\; {\rm d}\phi}
     {\cos^{n-3}\theta} \; .
\end{equation}
According to Fig.~2 and Fig.~3, $\Omega^{(3;n)}$ can be presented as 
a sum of six contributions:
\begin{eqnarray}
\Omega^{(3;n)} &=& 
\omega\left( {\textstyle{1\over2}}\varphi_{12}^{+}, \eta_{12} \right)
+ \omega\left( {\textstyle{1\over2}}\varphi_{12}^{-}, \eta_{12} \right)
\nonumber \\ &&
+ \omega\left( {\textstyle{1\over2}}\varphi_{23}^{+}, \eta_{23} \right)
+ \omega\left( {\textstyle{1\over2}}\varphi_{23}^{-}, \eta_{23} \right)
\nonumber \\ &&
+ \omega\left( {\textstyle{1\over2}}\varphi_{31}^{+}, \eta_{31} \right)
+ \omega\left( {\textstyle{1\over2}}\varphi_{31}^{-}, \eta_{31} \right),
\end{eqnarray}
with (see Refs.~\cite{Crete,DOS2,DK1})
\begin{equation}
\label{omega_int}
\omega\left( {\textstyle{1\over2}}\varphi, \eta \right)
= \frac{1}{2\varepsilon} \int\limits_0^{\varphi/2} {\rm d}\phi 
\left[ 1 - \left( 1+\frac{\tan^2\eta}{\cos^2\phi} \right)^{-\varepsilon}
\right] ,
\end{equation}
where $\varepsilon=\tfrac{1}{2}(4-n)$. 
Defining
$\tan\tfrac{\tau}{2}=\sin\eta \; \tan\tfrac{\varphi}{2}$,
we can obtain another useful representation, 
%for the integral in Eq.~(\ref{omega_int}),
\begin{eqnarray}
&& \hspace*{-7mm}
\int\limits_0^{\varphi/2} {\rm d}\phi 
\left( 1+\frac{\tan^2\eta}{\cos^2\phi} \right)^{-\varepsilon}
\nonumber \\ &&
= \sin\eta \; \cos^{2\varepsilon}\eta\; 
\int\limits_0^{\tau/2}
\frac{{\rm d}\psi\; \cos^{2\varepsilon}\psi}{1-\cos^2\eta\;\cos^2\psi}\; .
\end{eqnarray}

The remaining $\phi$-integral in Eq.~(\ref{omega_int}) 
can be calculated using a substitution
$\phi = \arctan\left(\frac{\sqrt{u}}{\sin\eta}\right)$.
The result can be presented in terms of Appell's hypergeometric
function $F_1$,
\begin{eqnarray}
\omega\left( {\textstyle{1\over2}}\varphi, \eta \right)
&=&\frac{1}{2\varepsilon}
\Bigl[ \frac{\varphi}{2}-\tan\tfrac{\varphi}{2}\; 
\cos^{2\varepsilon}\eta\;
\nonumber \\ &&
\hspace*{-15mm}
\times
F_1\left( {\textstyle{\frac{1}{2}}}, 1, \varepsilon ;
{\textstyle{\frac{3}{2}}} \bigl|~-\tan^2\tfrac{\varphi}{2},
-\tan^2\tfrac{\tau}{2} \right) \Bigr] .
\end{eqnarray} 
Moreover, using the transformation formula
\begin{eqnarray*}
F_1(a,b,b';c| x,y) &=& (1-x)^{-b} (1-y)^{-b'}\;
\\ &&
\hspace*{-15mm}
\times
F_1\left(c-a,b,b';c \biggl|~\frac{x}{x-1},\frac{y}{y-1}\right),
\end{eqnarray*}
the result can be presented as
\begin{eqnarray}
\label{omega_F1}
\omega\left( {\textstyle{1\over2}}\varphi, \eta \right)
&=&\frac{1}{2\varepsilon}
\Bigl[ \frac{\varphi}{2}- \sin\tfrac{\varphi}{2}\cos\tfrac{\varphi}{2}
\cos^{2\varepsilon}\tau_0\;
\nonumber \\ &&
\hspace*{-15mm}
\times
F_1\left( 1,1,\varepsilon; {\textstyle{\frac{3}{2}}}
\bigl|~\sin^2\tfrac{\varphi}{2},\; \sin^2\tfrac{\tau}{2} \right) \Bigr],
\end{eqnarray}
with $\cos\tau_0=\cos\eta\;\cos\frac{\tau}{2}$.
Similar $F_1$ functions occurred in Refs.~\cite{Tarasov-NPBPS,FJT}
(cf.\ also Ref.~\cite{CRSL}).

An important formula (shift $\varepsilon\to 1+\varepsilon$,
or $n\to n-2$) reads:
\begin{eqnarray}
\label{F1_recursion}
F_1\left( 1,1,\varepsilon; {\textstyle{\frac{3}{2}}}
\bigl|~x, y \right) 
= \frac{y}{x}\; _2F_1\left(  
1, 1+\varepsilon; {\textstyle{\frac{3}{2}}} \
\bigl|~y \right)
\nonumber \\ 
+ \left( 1-\frac{y}{x} \right)
F_1\left( 1,1,1\!+\!\varepsilon; {\textstyle{\frac{3}{2}}}
\bigl|~x, y \right)
\; .
\end{eqnarray}
It can be supplemented by a Kummer relation:
\begin{eqnarray}
\label{Kummer}
&& \hspace*{-7mm}
(1-2\varepsilon)\; _2F_1\left(
1, \varepsilon; \tfrac{3}{2}\bigl|~y \right)
\nonumber \\ &&
= 1 - 2\varepsilon (1-z)\;
_2F_1\left(
1, 1+\varepsilon; \tfrac{3}{2}\bigl|~y \right) \; .
\end{eqnarray}

Each of the three triangles in Fig.~2 may be associated with
a specific three-point function $J_i^{(3)}(n;1,1,1)$.
According to Eq.~(3.45) of~\cite{DD-JMP}, the result 
of such splitting reads
\begin{eqnarray}
\label{J3_split}
&& \hspace*{-7mm}
J^{(3)}(n;1,1,1) 
\nonumber \\ &&
= \frac{m_1^2 m_2^2 m_3^2}{\Lambda^{(3)}}\;
\sum_{i=1}^3 \frac{F_i^{(3)}}{m_i^2}\; J_i^{(3)}(n;1,1,1),
\end{eqnarray}
where 
$F_i^{(3)} = \frac{\partial}{\partial m_i^2} 
\left( m_i^2 \; D^{(3)}\right)$
(see also in Ref.~\cite{Nickel}).
The geometrical meaning of $F_i^{(3)}$ was discussed
in~\cite{DD-JMP}. In particular,  
\[
m_2^2 m_3^2\; F_1^{(3)}
+ m_1^2 m_3^2\; F_2^{(3)}
+ m_1^2 m_2^2\; F_3^{(3)}
= \Lambda^{(3)}
\]
means that the volume of the basic tetrahedron equals 
the sum of the volumes after splitting. 

For each of the integrals $J_i^{(3)}$, only one two-point
function appears in the reduction formulae. For instance,
using Eqs.~(\ref{omega_F1}) and (\ref{F1_recursion}) we get
\begin{eqnarray}
\label{J33}
&& \hspace*{-7mm}
(n-2) \pi^{-1}\; J_3^{(3)}(n+2;1,1,1) 
\nonumber \\ &&
= - 2 m_0^2\; J_3^{(3)}(n;1,1,1)
 - J^{(3)}(n;1,1,0) \; ,
\end{eqnarray}
and similarly for $J_1^{(3)}$ and $J_2^{(3)}$.
This yields a geometrical way to derive the recursion in $n$:
just take Eq.~(\ref{J3_split}), shift $n\to n+2$ 
and substitute Eq.~(\ref{J33}). The result is
\begin{eqnarray}
\label{dim_shift}
&& \hspace*{-8mm}
J^{(3)}(n+2;1,1,1) 
\nonumber \\ 
&\!=\!& -\frac{\pi m_1^2 m_2^2 m_3^2}{(n-2)\Lambda^{(3)}}
\biggl\{ 2 D^{(3)} J_3^{(3)}(n;1,1,1) 
\nonumber \\ && 
+ \frac{F_1^{(3)}}{m_1^2}\; J^{(3)}(n;0,1,1)
+ \frac{F_2^{(3)}}{m_2^2}\; J^{(3)}(n;1,0,1)
\nonumber \\ &&
+ \frac{F_3^{(3)}}{m_3^2}\; J^{(3)}(n;1,1,0) 
\biggr\} \; ,
\end{eqnarray}
in agreement with Refs.~\cite{Tarasov,FJT} 
(see also \cite{BDK,BGH+GG}).

\section{SPECIAL VALUES OF $n$}

\subsection{$n=3$ ($\varepsilon=\frac{1}{2}$)} 

In this case, we can use the known reduction formula 
for the $F_1$ function,
\begin{eqnarray}
&& \hspace*{-7mm}
F_1\big(a,b,b';b+b' \big|~x,y\big)
\nonumber \\ &&
= (1-y)^{-a}\; 
_2F_1\left( a, b; b+b' \bigl|~\frac{x-y}{1-y} \right) \; .
\end{eqnarray}
Taking into account that
\[
_2F_1\left( 1, 1; \tfrac{3}{2} \bigl|~z \right) 
= \frac{\arcsin\sqrt{z}}{\sqrt{z(1-z)}} \; ,
\]
we get
\[
F_1\left( 1, 1, {\textstyle{\frac{1}{2}}}; {\textstyle{\frac{3}{2}}}
\bigl|~\sin^2\tfrac{\varphi}{2},\; \sin^2\tfrac{\tau}{2} \right)
= \frac{\pi - 2\kappa}
  {\sin\varphi\; \cos\tau_0} \; ,
\]
with $\cos\kappa=\sin{\frac{\varphi}{2}}\cos\eta$ and
$\cos\tau_0=\cos{\frac{\tau}{2}}\cos\eta$.
Therefore,
\begin{equation}
\omega\left( {\textstyle{1\over2}}\varphi, \eta \right)\bigl|_{n=3}
= \frac{\varphi}{2} - \frac{\pi}{2} + \kappa \; .
\end{equation}
Collecting results for all six triangles, we reproduce Eq.~(5.4) 
of~\cite{DD-JMP} (see also in \cite{Nickel}),
\begin{equation}
\Omega^{(3;3)} = \Omega^{(3)} = \psi_{12}+\psi_{23}+\psi_{31}-\pi \; .
\end{equation}

\subsection{$n=2$ ($\varepsilon=1$)}

In this case, the function $F_1$ reduces to
\begin{eqnarray*}
&& \hspace*{-7mm}
F_1\left( 1, 1, 1; {\textstyle{\frac{3}{2}}}\bigl|~x,y \right)
\\
&\!=\!& \frac{1}{x\!-\!y}\left[
\frac{\sqrt{x}\;\arcsin{\sqrt{x}}}{\sqrt{1-x}} 
- \frac{\sqrt{y}\;\arcsin{\sqrt{y}}}{\sqrt{1-y}} 
\right] \; .
\end{eqnarray*}
In this way, we get
\[
F_1\left( 1, 1, 1; {\textstyle{\frac{3}{2}}}
\bigl|~\sin^2\tfrac{\varphi}{2}, \; \sin^2\tfrac{\tau}{2} \right)
= \frac{\varphi - \tau \sin\eta}
       {\sin\varphi \; \cos^2\tau_0}
\]
and, therefore,
\begin{equation}
\omega\left( {\textstyle{1\over2}}\varphi, \eta \right)\bigl|_{n=2}
= \tfrac{1}{4} \tau \sin{\eta} \; .
\end{equation}
Collecting results for all six triangles, we get 
\[
%\begin{equation}
\Omega^{(3;2)} = 
\tfrac{1}{2} \left(
\tau_{12}\sin\eta_{12}
+\tau_{23}\sin\eta_{23}
+\tau_{13}\sin\eta_{13}
\right)
\; .
\]
Recalling that the two-point integral in two dimensions is proportional
to $\tau/\sin\tau$ (see Eq.~(4.3) of Ref.~\cite{DD-JMP}), we see that 
the three-point integral (with $n=2$) is 
a combination of three two-point integrals, with coefficients 
proportional to $\sin\tau_{jl}\sin\eta_{jl}$ 
(cf.\ Eq.~(10) of Ref.~\cite{Nickel}).

\subsection{$n=5$ ($\varepsilon=-{\textstyle{\frac{1}{2}}}$)}

In this case, we obtain
\begin{eqnarray*}
\omega\left( {\textstyle{1\over2}}\varphi, \eta \right)\bigl|_{n=5}
&=& \frac{\pi}{2}  -\frac{\varphi}{2} - \kappa 
\nonumber \\ &&
+ \frac{1}{2}\tan\eta\; 
\ln\left(\frac{1+\sin\frac{\tau}{2}}{1-\sin\frac{\tau}{2}}\right) .
\end{eqnarray*}
Collecting results for all six triangles, we get
\begin{eqnarray*}
\Omega^{(3;5)} &=& -\left( \psi_{12}+\psi_{23}+\psi_{31}-\pi \right)
\nonumber \\ &&
+\tan\eta_{12}\; 
\ln\left(\frac{m_1+m_2+\sqrt{k_{12}^2}}{m_1+m_2-\sqrt{k_{12}^2}}\right)
\nonumber \\ &&
+\tan\eta_{23}\;
\ln\left(\frac{m_2+m_3+\sqrt{k_{23}^2}}{m_2+m_3-\sqrt{k_{23}^2}}\right)
\nonumber \\ &&
+\tan\eta_{13}\;
\ln\left(\frac{m_1+m_3+\sqrt{k_{13}^2}}{m_1+m_3-\sqrt{k_{13}^2}}\right) .
\end{eqnarray*}
In other words, the five-dimensional three-point integral can be
expressed in terms of the three-dimensional three- 
and two-point integrals (see Eqs.~(5.4) and (4.6) of~\cite{DD-JMP}),
in agreement with Eq.~(\ref{dim_shift}).

\subsection{$n=4$ ($\varepsilon\to 0$)}

In this case, we need to expand the $F_1$ function up to the term
linear in $\varepsilon$ (it is easy to see that the $\varepsilon^0$ 
term cancels the $\frac{\varphi}{2}$ contribution).
The result can be presented in terms of Clausen function,
see Eq.~(5.21) of~\cite{DD-JMP}. Collecting three contributions
of this type, we get $6\times 3=18$ Clausen functions
that can be analytically continued in terms of 
12 dilogarithms~\cite{tHV79}.
In Ref.~\cite{Wagner} the result is presented in terms 
of 15 Clausen functions. 

\section{ANALYTIC CONTINUATION}

In the integral occurring in Eq.~(\ref{omega_int}),
let us substitute
$z \Rightarrow e^{2{\rm i}\phi}$,
so that
$\cos^2\phi \Rightarrow \frac{(1+z)^2}{4z}$ and 
\begin{equation}
1 + \frac{\tan^2\eta}{\cos^2\phi} \Rightarrow 
\frac{(z+\rho)(z+1/\rho)}{(z+1)^2}, 
\end{equation}
with
\begin{equation}
\rho \equiv \frac{1-\sin\eta}{1+\sin\eta} \; .
\end{equation}
In this way, we get
\begin{eqnarray}
&& \hspace*{-7mm} 
\int\limits_0^{\varphi/2} {\rm d}\phi\; 
\left( 1 + \frac{\tan^2\eta}{\cos^2\phi} \right)^{-\varepsilon} 
\nonumber \\ &&
\Rightarrow
\frac{{\rm i}}{2}\;
\int\limits_{z_0}^1 \frac{{\rm d}z}{z}\;
\left[ \frac{(z+\rho)(z+1/\rho)}{(z+1)^2} \right]^{-\varepsilon}, 
\end{eqnarray}
with $z_0\leftrightarrow e^{{\rm i}\varphi}$.
Expanding in $\varepsilon$, we get 
\begin{equation}
\label{Qj}
Q_j \equiv \int\limits_{z_0}^1 \frac{{\rm d}z}{z}\;
\ln^j\left[ \frac{(z+\rho)(z+1/\rho)}{(z+1)^2} \right] \; .
\end{equation}

The first term,
\begin{eqnarray}
\label{Q1}
Q_1 
&=& \Li{2}{-z_0\rho} + \Li{2}{-z_0/\rho} - 2\Li{2}{-z_0}
\nonumber \\ &&
+ \tfrac{1}{2}\ln^2\rho \; ,
\end{eqnarray} 
yields the known result~\cite{tHV79} for the three-point function in four
dimensions.
The r.h.s.\ of Eq.~(\ref{Q1}) can also be presented as
\begin{eqnarray}
&& \hspace*{-7mm}
2\Li{2}{z_0} - \Li{2}{\frac{\rho+z_0}{\rho+z_0^{-1}}}
- \Li{2}{z_0^2\;\frac{\rho+z_0^{-1}}{\rho+z_0}}
\nonumber \\ &&
+ \tfrac{1}{2}\ln^2\rho
- \tfrac{1}{2}\ln^2\left[\frac{z_0\rho(\rho+z_0^{-1})}{\rho+z_0}\right]
\; , \hspace*{7mm}
\end{eqnarray}
in agreement with Eq.~(5.17) of Ref.~\cite{DD-JMP}. 

The second term, $Q_2$, gives the $\ep$ term of the three-point
function. For $j=2$, the integral~(\ref{Qj}) can be evaluated in terms 
of polylogarithms,
\begin{equation}
Q_2 = Q_2^{(1)}(z_0, \rho) \ln\rho + Q_2^{(2)}(z_0, \rho) \; ,
\end{equation}
\begin{eqnarray*}
Q_2^{(1)} 
&\!\!=\!\!& 
2\Li{2}{\frac{1-\rho}{1+z_0\rho}}
+2\Li{2}{\frac{z_0(\rho-1)}{1+z_0\rho}}
\nonumber \\ &&
-2\Li{2}{\frac{\rho-1}{z_0+\rho}}
-2\Li{2}{\frac{z_0(1-\rho)}{z_0+\rho}}
\nonumber \\ &&
-\Li{2}{\frac{1-\rho^2}{1+z_0\rho}}
-\Li{2}{\frac{z_0(\rho^2-1)}{\rho(1+z_0\rho)}}
\nonumber \\ &&
+\Li{2}{\frac{\rho^2-1}{\rho(z_0\!+\!\rho)}}
+\Li{2}{\frac{z_0(1\!-\!\rho^2)}{z_0+\rho}},
\\
Q_2^{(2)} &\!\!=\!\!&
4\Snp{1,2}{\frac{1-\rho}{1+z_0\rho}}
-4\Snp{1,2}{\frac{z_0(\rho-1)}{1+z_0\rho}}
\nonumber \\ &&
+4\Snp{1,2}{\frac{\rho-1}{z_0+\rho}}
-4\Snp{1,2}{\frac{z_0(1-\rho)}{z_0+\rho}}
\nonumber \\ &&
-\Snp{1,2}{\frac{1-\rho^2}{1+z_0\rho}}
+\Snp{1,2}{\frac{z_0(\rho^2-1)}{\rho(1+z_0\rho)}}
\nonumber \\ &&
-\Snp{1,2}{\frac{\rho^2-1}{\rho(z_0\!+\!\rho)}}
+\Snp{1,2}{\frac{z_0(1\!-\!\rho^2)}{z_0+\rho}} .
\end{eqnarray*}
Note that the arguments of 
Nielsen polylogarithms ${\rm S}_{1,2}$ are the same
as the arguments of ${\rm Li}_2$. Furthermore,
\begin{eqnarray*}
\frac{\partial}{\partial\rho} \; Q_2^{(2)}(z_0,\rho)
&=& - \ln\rho\; \frac{\partial}{\partial\rho}\; Q_2^{(1)}(z_0,\rho) \; , 
\\
\rho\;\frac{\partial}{\partial\rho}\; Q_2(z_0,\rho) 
&=& Q_2^{(1)}(z_0,\rho) \; .
\end{eqnarray*}
Another useful representation for $Q_2^{(1)}$ reads
\begin{eqnarray}
Q_2^{(1)}
&\!\!=\!\!& 2\Li{2}{-\frac{\rho+z_0}{1+\rho z_0}}
-  2\Li{2}{-\frac{1+\rho z_0}{\rho+z_0}}
\nonumber \\ &&
+ 2 \ln\left[ \frac{\rho}{(\rho+1)^2}\right]
\ln\left( \frac{1+\rho z_0}{\rho+z_0} \right) .
\end{eqnarray}

The occurring ${\rm S}_{1,2}$ functions can be
presented in terms of trilogarithms ${\rm Li}_3$.
This result corresponds to Eq.~(82) of Ref.~\cite{FJT}.
We note that the first calculation of the $\ep$-term of
the one-loop three-point function was given in Ref.~\cite{NMB}
(see also Ref.~\cite{UD+DT} for the off-shell massless case).

Eq.~(\ref{Qj}) shows that all higher terms of the $\ep$-expansion
of the one-loop three-point function can be expressed 
in terms of one-fold integrals of the products of 
logarithms of three linear arguments. 
(For some specific configurations, the $\ep^2$ terms were studied
in Ref.~\cite{KMR}.) The considered representations
may be useful to understand the types of generalized functions
needed to decsribe the analytic structure of the results 
for higher terms of the $\ep$-expansion. 

\noindent
{\bf Acknowledgements.} I am thankful to
R.~Delbourgo, M.Yu.~Kalmykov, Z.~Merebashvili, and O.V.~Tarasov 
for useful discussions. I am grateful to
the organizers of ACAT05 for their hospitality.

\end{document}